# Acoustic Impedance Calculation via Numerical Solution of the Inverse Helmholtz Problem

**Danish Patel (Corresponding author)**
School of Mechanical Engineering
Purdue University
585 Purdue Mall
West Lafayette IN 47906, USA
patel472@purdue.edu
Ph: +1-765-479-5099

**Prateek Gupta**
School of Mechanical Engineering
Purdue University
585 Purdue Mall
West Lafayette IN 47906, USA
gupta288@purdue.edu
Ph: +1-765-409-9187

**Carlo Scalo**
School of Mechanical Engineering
Purdue University
585 Purdue Mall
West Lafayette IN 47906, USA
scalo@purdue.edu
Ph: +1-765-775-3153

**Abstract**

Assigning homogeneous boundary conditions, such as acoustic impedance, to the thermoviscous wave equations (TWE) derived by transforming the linearized Navier-Stokes equations (LNSE) to the frequency domain yields a so-called Helmholtz solver, whose output is a discrete set of complex eigenfunction and eigenvalue pairs. The proposed method – the inverse Helmholtz solver (iHS) – reverses such procedure by returning the value of acoustic impedance at one or more unknown impedance boundaries (IBs) of a given domain via spatial integration of the TWE for a given real-valued frequency with assigned conditions on other boundaries. The iHS procedure is applied to a second-order spatial discretization of the TWEs derived on an unstructured grid with staggered grid arrangement. The momentum equation only is extended to the center of each IB face where pressure and velocity components are co-located and treated as unknowns. One closure condition considered for the iHS is the assignment of the surface gradient of pressure phase over the IBs, corresponding to assigning the shape of the acoustic waveform at the IB. The iHS procedure is carried out independently for each frequency in order to return the complete broadband complex impedance distribution at the IBs in any desired frequency range. The iHS approach is first validated against Rott's theory for both inviscid and viscous, rectangular and circular ducts where the solver's capability to return the value of tangential impedance is also assessed. The impedance of a geometrically complex toy porous cavity is then reconstructed and verified against companion full compressible unstructured Navier-Stokes simulations resolving the cavity geometry and one-dimensional impedance test tube calculations based on time-domain impedance boundary conditions (TDIBC). The iHS methodology is also shown to capture thermoacoustic effects, with reconstructed impedance values quantitatively in agreement with thermoacoustic growth rates.

## 1. Introduction

Acoustic impedance in a compressible flow is defined as [1] the ratio of the Fourier-transformed pressure to velocity fluctuations evaluated at a given point in space, $\mathbf{x} = \{x_1, x_2, x_3\}$, and angular frequency, $\omega$. Hereafter, the following convention will be adopted for harmonic pressure and velocity fluctuations, respectively,

$$p'(\mathbf{x},t) = \text{Re}\left(\hat{p}(\mathbf{x};\omega)\,e^{j\omega t}\right), \quad \mathbf{u}'(\mathbf{x},t) = \text{Re}\left(\hat{\mathbf{u}}(\mathbf{x};\omega)\,e^{j\omega t}\right), \quad (1)$$

where the superscript $(')$ indicates fluctuations about a base state and $(\hat{\ })$ their respective complex amplitudes, with $j = \sqrt{-1}$ being the imaginary unit. The same convention is applied to the rest of the fluctuating quantities as discussed in §2. An acoustic impedance can be associated to each velocity component, $\hat{\mathbf{u}}(\mathbf{x};\omega) = \{\hat{u}_1(\mathbf{x};\omega), \hat{u}_2(\mathbf{x};\omega), \hat{u}_3(\mathbf{x};\omega)\}$, hence yielding the definition

$$Z_i(\mathbf{x};\omega) \equiv \frac{\hat{p}(\mathbf{x};\omega)}{\hat{u}_i(\mathbf{x};\omega)}, \qquad i = 1, 2, 3. \quad (2)$$

The acoustic admittance is defined as the reciprocal of the impedance, $Y_i(\mathbf{x};\omega) = Z_i(\mathbf{x};\omega)^{-1}$, and can also be represented in vector notation, $\mathbf{Y}(\mathbf{x};\omega) \equiv \hat{\mathbf{u}}(\mathbf{x};\omega)/\hat{p}(\mathbf{x};\omega)$. More commonly, impedance is defined based on the velocity component taken along the normal $\hat{\mathbf{n}}$ to a given surface which will be referred to as impedance boundary (IB), that is

$$Z_n(\mathbf{x};\omega) = \frac{\hat{p}(\mathbf{x};\omega)}{\hat{u}_n(\mathbf{x};\omega)}, \qquad \forall \ \mathbf{x} \in \text{IB} \quad (3)$$

where $u_n(\mathbf{x};\omega) = \hat{\mathbf{u}}(\mathbf{x};\omega) \cdot \hat{\mathbf{n}}(\mathbf{x})$, and will be hereafter referred to as normal impedance, related to normal admittance via $Y_n = Z_n^{-1}$.

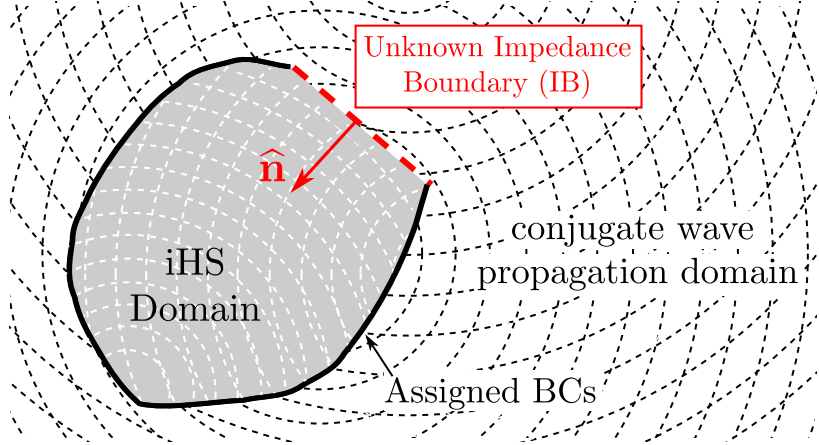

Figure 1: Illustration of a wave propagation problem divided into an inverse Helmholtz Solver (iHS) domain and a conjugate wave propagation domain. The iHS can provide the broadband impedance at the IB, which can be used as an impedance boundary condition in a simulation solving the conjugate wave propagation domain only.

The impedance is a complex quantity, whose real and imaginary parts are referred to as resistance, $R(\mathbf{x}; \omega) = \mathrm{Re}\left[Z(\mathbf{x}; \omega)\right]$, and reactance, $X(\mathbf{x}; \omega) = \mathrm{Im}\left[Z(\mathbf{x}; \omega)\right]$, respectively. The admittance can similarly be decomposed into its real and imaginary parts called conductance, $G(\mathbf{x}; \omega) = \mathrm{Re}[Y(\mathbf{x}; \omega)]$, and susceptance, $B(\mathbf{x}; \omega) = \mathrm{Im}[Y(\mathbf{x}; \omega)]$, respectively. The base impedance, defined as $Z_0 = \rho_0 a_0$, where $\rho_0$ and $a_0$ are the base density and speed of sound, is often used to obtain the specific (or normalized) values of acoustic impedance and admittance, hereafter referred as $Z_* = Z/\rho_0 a_0$ and $Y_* = \rho_0 a_0 Y$, respectively. Unless otherwise stated, the terms impedance, $Z$ and admittance, $Y$ will refer to normal impedance and normal admittance respectively in the remainder of this manuscript.

This manuscript presents a novel computational methodology—namely, the inverse Helmholtz solver (iHS), capable of deriving the full broadband impedance (2) on a given impedance boundary (IB) that separates two domains: (i) the isolable iHS domain, and (ii) the conjugate wave propagation domain as has been illustrated in figure 1. The iHS is limited to a linear assumption and therefore non-linear effects, where dominant, that would otherwise impact the impedance at the IB are neglected. Despite the linear assumption, it is algebraically shown that some information concerning the wave propagation in the conjugate domain (figure 1) is needed for closure of the iHS problem (§2.4). Finally, we point out that no assumptions on the isentropic nature of the flow are made in the formulation of the iHS, despite the focus being primarily on velocity and pressure fluctuation variables (see §2.6).

The acoustic properties of a system are typically investigated via Helmholtz solvers, which rely on an eigenvalue formulation (12), hence yielding a discrete set of complex eigenvalues and eigenvectors containing frequency and growth rate, and resonant waveform information respectively. Such approach requires homogeneous boundary conditions to be assigned everywhere (figure 3a). In an inviscid problem, for example, only the normal impedance (3) needs to be specified at all boundaries to close the eigenvalue problem. In a viscous problem, additional boundary conditions, related to the tangential components of velocity, and temperature, need to be specified: they can be in the form of homogeneous Dirichlet or Neumann conditions such as slip/no-slip for the velocity, and isothermal/adiabatic for the temperature fluctuations, or, for example, be expressed as ratios of the frequency-transformed pressure to the tangential components of velocity (tangential impedance) or temperature. A classic eigenvalue approach therefore cannot be used to *calculate* the spatial distribution of the broadband impedance, for example, at the orifice of an acoustically absorptive cavity or the inlet of a nozzle. A priori knowledge of impedance at these locations (IB) would yield boundary conditions for simulations to be carried out only on the conjugate wave propagation side of the impedance surface (e.g. high-fidelity time-domain simulation of the flow grazing such cavity or of the reactive flow in the combustion chamber upstream of the nozzle), sparing the need to resolve the flow on the other side of the same surface (e.g. in the cavity or in the nozzle respectively), which, along with the IB, would form the iHS domain.

The iHS methodology presented herein does *not* belong to the class of inverse problems in acoustics that are concerned with the reconstruction of one or multiple acoustic sources from far field acoustic measurements [2, 3]. It is rather more appropriately classifiable as an inverse Cauchy problem [4] applied to the thermoviscous-wave equations (TWE): the unknown, frequency-transformed solution at one or multiple boundaries is found by assigning the angular frequency $\omega$ (as per ansatz (1)) and the phase distribution of either velocity or pressure fluctuations at said boundary with the solution being known on the other boundaries (and not necessarily homogeneous). Such problems are typically solved using iterative techniques [5, 6] and rely on an initial guess for the unknown boundary values that is updated until a given criterion is satisfied within a prescribed tolerance. Such methods are computationally expensive since they mandate the solution of a direct problem at each step, and may often not be able to satisfy the convergence criterion to the desired accuracy [7]. The iHS methodology solves the inverse Cauchy problem applied to the linearized compressible Navier-Stokes equations via a one-time direct solution of a system of linear equations.

Inverse problems associated with acoustic source reconstruction or the Helmholtz equation do not typically have a unique or well-defined solution, and may require the adoption of regularization techniques [8, 9]. Grayzin et al. [5] applied Sommerfeld radiation conditions to the Helmholtz problem thus obtaining an overconstrained linear system of equations, which was then solved using least squares. A meshless, direct method to determine unknown boundary conditions to the Helmholtz problem was proposed by Jin & Zheng [10] relying on the method of fundamental solutions and truncated singular value decomposition regularization. Piechowicz et al. [11, 12, 13] have developed a numerical scheme that relies on a boundary element method applied to the Helmholtz-Kirchoff equation to evaluate the acoustic impedance of obstacles and/or boundaries internal to a closed domain from pressure information taken at a finite set of interior points. To the best of authors' knowledge no prior work has tackled the evaluation of the spatial distribution of the broadband acoustic impedance (2) over one or multiple boundaries of a computational domain without the use of regularization or iterative techniques.

The iHS methodology is outlined in section §2 of the present manuscript, where the linearized governing equations are first presented for a general compressible fluid (§2.1), yielding the set of TWEs, which are then cast in the form of an inverse Helmholtz problem with appropriate closure conditions (§2.3). The iHS is first validated against Rott's thermoviscous theory [14, 15] in rectangular and circular constant cross-sectional area ducts, with both inviscid and viscous wave propagation assumption in a quiescent fluid with a uniform thermodynamic base state (section §3). The evaluation of the impedance of thermoacoustically active cavities is also demonstrated, quantitatively predicting the growth rate of unstable modes (§4). The impedance of a geometrically complex acoustically absorptive toy cavity is then considered (section §5), validated against time-domain pore-resolved unstructured Navier-Stokes simulations under both harmonic (§5.2) and pulsed (§5.3) excitation, and against simulations based on time-domain impedance boundary conditions (TDIBC) [16, 17, 18].

## 2. Formulation of the inverse Helmholtz solver

In this section, the inverse Helmholtz problem formulation (§2.3) is derived starting from the linearized compressible Navier-Stokes equations (§2.1) discretized on an unstructured grid with a staggered variable arrangement (§2.2).

### 2.1. Linearized Navier-Stokes Equations (LNSE)

Governing equations for a fully compressible flow are

$$\frac{\partial}{\partial t}(\rho) + \frac{\partial}{\partial x_k}(\rho u_k) = 0, \tag{4a}$$

$$\frac{\partial}{\partial t}(\rho u_i) + \frac{\partial}{\partial x_k}(\rho u_i u_k) = -\frac{\partial}{\partial x_i}p + \frac{\partial}{\partial x_k}\tau_{ik}, \tag{4b}$$

$$\frac{\partial}{\partial t}(\rho e) + \frac{\partial}{\partial x_k}[u_k(\rho e + p)] = \frac{\partial}{\partial x_k}(u_i \tau_{ik} - q_k), \tag{4c}$$

where $x_1$, $x_2$, and $x_3$ (or $x$, $y$, and $z$) are Cartesian coordinates, $u_i$ are the velocity field components in each of these directions, and $p$, $\rho$, and $e$ are, respectively, the pressure, density, and total energy per unit mass.

The viscous shear stress, $\tau_{ik}$ and heat flux $q_k$ are defined as,

$$\tau_{ik} = 2\mu \left[ \frac{1}{2} \left( \frac{\partial u_k}{\partial x_i} + \frac{\partial u_i}{\partial x_k} \right) \right] + \lambda \frac{\partial u_m}{\partial x_m} \delta_{ik}, \qquad q_k = -\kappa \frac{\partial T}{\partial x_k},$$

where $\mu$ and $\lambda$ are the first and second viscosity coefficients, respectively, $\kappa$ is the thermal conductivity, and $T$ is the temperature.

Decomposing all variables in (4) into a base state, denoted with the subscript (0), and a fluctuation, denoted with the superscript (') and linearizing, yields

$$\frac{\partial \rho'}{\partial t} + \frac{\partial (\rho_0 u'_k)}{\partial x_k} + \frac{\partial (u_{0,k} \rho')}{\partial x_k} = 0, \tag{5a}$$

$$\rho_0 \frac{\partial u'_i}{\partial t} + u_{0,i} \frac{\partial \rho'}{\partial t} + \frac{\partial (\rho_0 u'_i u_{0,k})}{\partial x_k} + \frac{\partial (\rho' u_{0,i} u_{0,k})}{\partial x_k} + \frac{\partial (\rho_0 u_{0,i} u'_k)}{\partial x_k} = -\frac{\partial p'}{\partial x_i} + \frac{\partial}{\partial x_k} \tau'_{ik}, \tag{5b}$$

$$\begin{aligned}
\rho_0 \frac{\partial e'}{\partial t} + e_0 \frac{\partial \rho'}{\partial t} &+ \frac{\partial (\rho' e_0 u_{0,k})}{\partial x_k} + \frac{\partial (\rho_0 e' u_{0,k})}{\partial x_k} + \frac{\partial (\rho_0 e_0 u'_k)}{\partial x_k} + p_0 \frac{\partial u'_k}{\partial x_k} \\
&+ p' \frac{\partial u_{0,k}}{\partial x_k} - \kappa \frac{\partial}{\partial x_k} \left( \frac{\partial}{\partial x_k} T' \right) = 2\lambda \frac{\partial u_{0,j}}{\partial x_j} \frac{\partial u'_k}{\partial x_k} + 2\mu \frac{\partial u_{0,k}}{\partial x_j} \frac{\partial u'_k}{\partial x_j} + 2\mu \frac{\partial u_{0,j}}{\partial x_k} \frac{\partial u'_k}{\partial x_j}.
\end{aligned} \tag{5c}$$

The second viscosity coefficient, $\lambda$, accounts for the bulk viscous dissipation and is related to the bulk viscosity via $\mu_B \equiv \lambda + (2/3)\mu$. A frequency dependent formulation for bulk viscosity, as proposed by Lin *et al.* [19], can be conveniently adopted in the frequency-transformed version of the LNSE, discussed below. Note that the second viscosity has been denoted by $\mu'$ in [19] and should not be confused with fluctuations in dynamic viscosity, $\mu$, which are always neglected. Unless otherwise specified, bulk viscosity will be assumed to be negligible. Fluctuations in other fluid properties such as $\kappa$ and $\lambda$ are also neglected.

The total differentials of pressure and internal energy read,

$$dp = \frac{1}{\rho \beta_T} d\rho + \frac{\alpha_v}{\beta_T} dT, \tag{6}$$

$$de = c_v dT - B_0 d\rho \tag{7}$$

with,

$$\alpha_v = -\left. \rho \frac{\partial p}{\partial T} \right|_v \left( \left. \frac{\partial p}{\partial v} \right|_T \right)^{-1}, \quad \beta_T = -\left( \left. v \frac{\partial p}{\partial v} \right|_T \right)^{-1}, \quad B_0 = \frac{C_0}{\rho^2}, \quad \text{and} \quad C_0 = \left( T \left. \frac{\partial p}{\partial T} \right|_v - p \right)$$

where, the coefficient of thermal expansion, $\alpha_v$, the isothermal compressibility, $\beta_T$, and constants $B_0$ and $C_0$ can be obtained from the equation of state. Substituting (6) and (7) in the system of equations (5), and eliminating density perturbation yields the linearized governing equations for mass

$$\begin{aligned}
\rho_0 \beta_T \frac{\partial p'}{\partial t} - \rho_0 \alpha_v \frac{\partial T'}{\partial t} + \rho_0 \frac{\partial u'_k}{\partial x_k} + u'_k \frac{\partial \rho_0}{\partial x_k} + \rho_0 \beta_T u_{0,k} \frac{\partial p'}{\partial x_k} + p' u_{0,k} \frac{\partial (\rho_0 \beta_T)}{\partial x_k} \\
- \rho_0 \alpha_v u_{0,k} \frac{\partial T'}{\partial x_k} - T' u_{0,k} \frac{\partial (\rho_0 \alpha_v)}{\partial x_k} + \rho_0 p' \beta_T \frac{\partial u_{0,k}}{\partial x_k} - \rho_0 T' \alpha_v \frac{\partial u_{0,k}}{\partial x_k} = 0,
\end{aligned} \tag{8}$$

momentum

$$\begin{aligned}
\rho_0 \frac{\partial u'_i}{\partial t} + \rho_0 u'_i \frac{\partial u_{0,k}}{\partial x_k} + \rho_0 u_{0,k} \frac{\partial u'_i}{\partial x_k} + u'_i u_{0,k} \frac{\partial \rho_0}{\partial x_k} + \rho_0 \beta_T p' u_{0,k} \frac{\partial u_{0,i}}{\partial x_k} - \rho_0 \alpha_v T' u_{0,k} \frac{\partial u_{0,i}}{\partial x_k} + \rho_0 u'_k \frac{\partial u_{0,i}}{\partial x_k} + \frac{\partial p'}{\partial x_i} \\
= \mu \frac{\partial}{\partial x_k} \left( \frac{\partial}{\partial x_k} u'_i \right) + (\mu + \lambda) \frac{\partial}{\partial x_i} \left( \frac{\partial u'_k}{\partial x_k} \right),
\end{aligned} \tag{9}$$

and energy

$$(\rho_0 c_v + C_0 \alpha_v) \frac{\partial T'}{\partial t} - C_0 \beta_T \frac{\partial p'}{\partial t} + p' \rho_0 \beta_T u_{0,k} \frac{\partial e_0}{\partial x_k} - T' \rho_0 \alpha_v u_{0,k} \frac{\partial e_0}{\partial x_k} + \rho_0 u'_k \frac{\partial e_0}{\partial x_k} + c_v T' \frac{\partial (\rho_0 u_{0,k})}{\partial x_k}$$
$$- \frac{C_0}{\rho_0} \beta_T p' \frac{\partial (\rho_0 u_{0,k})}{\partial x_k} + \frac{C_0}{\rho_0} \alpha_v T' \frac{\partial (\rho_0 u_{0,k})}{\partial x_k} + \rho_0 c_v u_{0,k} \frac{\partial T'}{\partial x_k} + \rho_0 T' u_{0,k} \frac{\partial c_v}{\partial x_k} - C_0 \beta_T u_{0,k} \frac{\partial p'}{\partial x_k}$$
$$- p' \frac{C_0}{\rho_0} u_{0,k} \frac{\partial (\rho_0 \beta_T)}{\partial x_k} + C_0 \alpha_v u_{0,k} \frac{\partial T'}{\partial x_k} + T' \frac{C_0}{\rho_0} u_{0,k} \frac{\partial (\rho_0 \alpha_v)}{\partial x_k} - p' \rho_0^2 \beta_T u_{0,k} \frac{\partial B_0}{\partial x_k} + T' \rho_0^2 \alpha_v u_{0,k} \frac{\partial B_0}{\partial x_k} \quad (10)$$
$$+ p_0 \frac{\partial u'_k}{\partial x_k} + p' \frac{\partial u_{0,k}}{\partial x_k} - \kappa \frac{\partial}{\partial x_k} \left( \frac{\partial}{\partial x_k} T' \right) = 2\lambda \frac{\partial u_{0,j}}{\partial x_j} \frac{\partial u'_k}{\partial x_k} + 2\mu \frac{\partial u_{0,k}}{\partial x_j} \frac{\partial u'_k}{\partial x_j} + 2\mu \frac{\partial u_{0,j}}{\partial x_k} \frac{\partial u'_k}{\partial x_j}.$$

*2.1.1. Ideal gas and quiescent flow approximations*

Unless otherwise specified, the results shown in this manuscript are generated assuming a quiescent ($u_{0,i} = 0$) but not necessarily uniform base state, in an ideal gas ($\alpha_v = 1/T_0$, $\beta_T = 1/p_0$, $B_0 = C_0 = 0$), resulting in equations (8), (9), and (10) to be simplified to

$$\rho_0 \beta_T \frac{\partial p'}{\partial t} - \rho_0 \alpha_v \frac{\partial T'}{\partial t} + \rho_0 \frac{\partial u'_k}{\partial x_k} + u'_k \frac{\partial \rho_0}{\partial x_k} = 0, \quad (11a)$$

$$\rho_0 \frac{\partial u'_i}{\partial t} + \frac{\partial p'}{\partial x_i} - \mu \frac{\partial}{\partial x_k} \left( \frac{\partial}{\partial x_k} u'_i \right) - (\mu + \lambda) \frac{\partial}{\partial x_i} \left( \frac{\partial u'_k}{\partial x_k} \right) = 0, \quad (11b)$$

$$\rho_0 c_v \frac{\partial T'}{\partial t} + \rho_0 c_v u'_k \frac{\partial T_0}{\partial x_k} + p_0 \frac{\partial u'_k}{\partial x_k} - \kappa \frac{\partial}{\partial x_k} \left( \frac{\partial}{\partial x_k} T' \right) = 0, \quad (11c)$$

respectively.

*2.2. Thermoviscous wave equations and numerical discretization*

Transforming the equations (8 -10) from time domain to frequency domain according to the convention (1), yields the thermoviscous wave equations, or TWE, not reported here for brevity. The TWE are discretized via a second-order spatial discretization and staggered variable arrangement (figure 2) collocating thermodynamic field variables (pressure $\hat{p}$, specific energy $\hat{e}$, and hence temperature $\hat{T}$ and density $\hat{\rho}$) at the mesh element centroids and velocity field components $\hat{u}_i$ at the face centers.

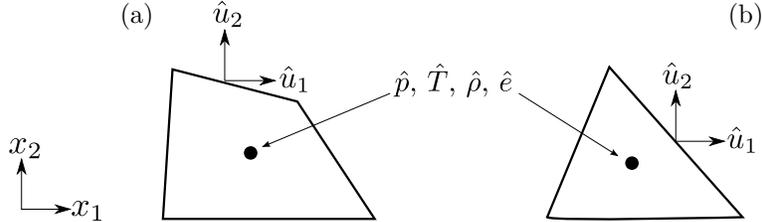

Figure 2: Staggered variable arrangement over a quadrilateral (a), and triangular (b) cell. Thermodynamic quantities, $p$, $T$, $\rho$, $e$ and $c_v$, are stored at the cell centres ($\bullet$), while velocity components, $u_i$, are located on face centres.

Upon applying a spatial discretization and homogeneous boundary conditions, the TWE read

$$\mathrm{j}\,\sigma \underline{X} + \underline{\underline{A}}_{NS} \underline{X} = 0 \quad (12)$$

where $\underline{X} = \left\{ \underline{\hat{p}}, \underline{\hat{T}}, \underline{\hat{u}}_1, \underline{\hat{u}}_2, \underline{\hat{u}}_3 \right\}^T$. For a generic fluid-thermo-dynamic quantity $\hat{\phi}$, $\underline{\hat{\phi}}$ represents the column array collecting all discretized values of the complex amplitude of the fluctuations $\hat{\phi}$. Equation (12) describes a conventional eigenvalue problem yielding a discrete spectrum of frequencies and growth rates (eigenvalues), and wave forms (eigenvectors) of the problem, as conceptually illustrated in figure 3a.

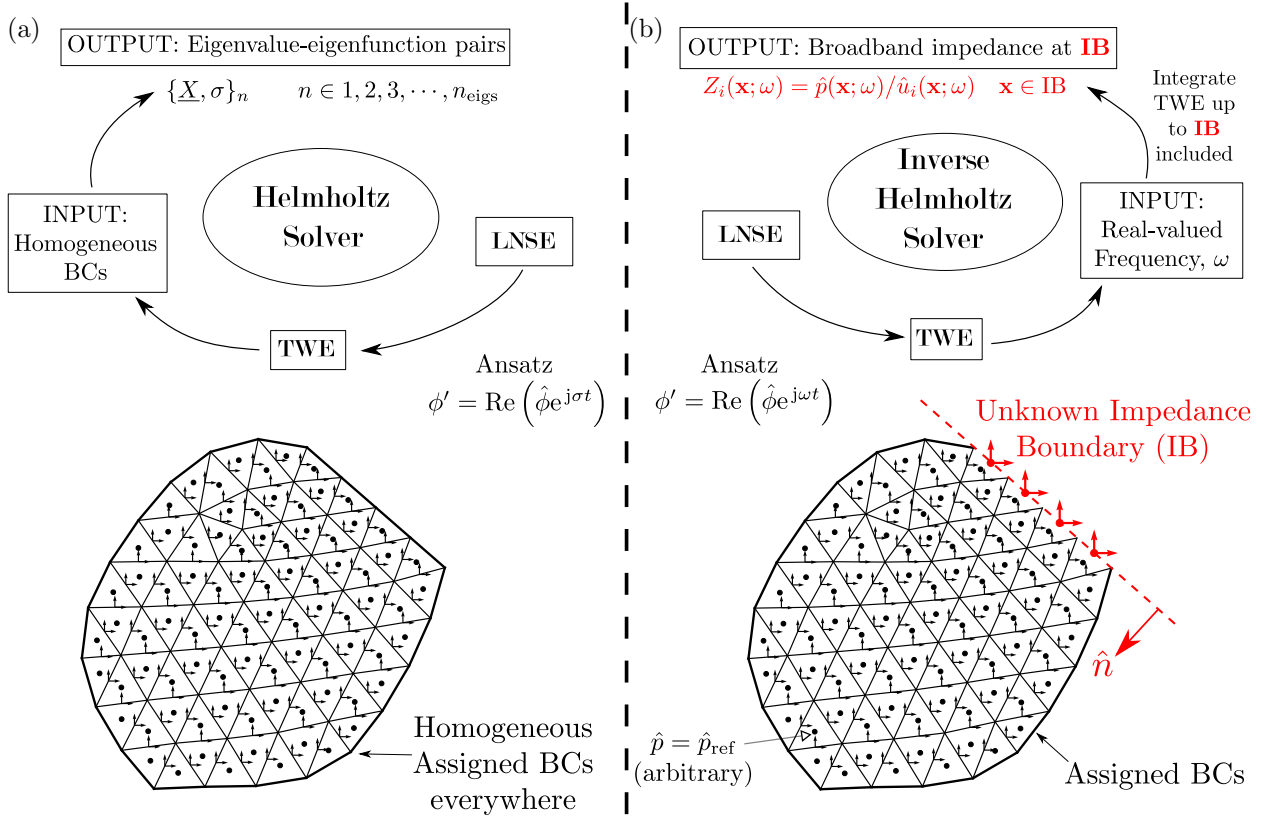

Figure 3: Conceptual illustration of eigenvalue-based or direct (a) and inverse (b) Helmholtz solvers. Both approaches are based on the linearized Navier-Stokes equations (LNSE) i.e. equations (8 -10) and their Fourier transformed equivalents – the thermoviscous wave equations (TWE).

## 2.3. Inverse Helmholtz problem formulation

To derive the inverse Helmholtz solver (iHS) formulation, the TWE are, rather, considered for an assigned real-valued frequency $\omega$, treated as an input parameter and extended to the IB where four new unknowns $(\{\hat{\underline{u}}_{1,\text{IB}}, \hat{\underline{u}}_{2,\text{IB}}, \hat{\underline{u}}_{3,\text{IB}}\}^T, \hat{p}_{\text{IB}})$ are introduced (shown in red in figure 3b). The iHS system reads:

$$\mathrm{j}\omega \begin{pmatrix} \hat{\underline{p}} \\ \hat{\underline{\tilde{T}}} \\ \hat{\underline{u}}_1 \\ \hat{\underline{u}}_2 \\ \hat{\underline{u}}_3 \\ \hat{\underline{u}}_{1,\text{IB}} \\ \hat{\underline{u}}_{2,\text{IB}} \\ \hat{\underline{u}}_{3,\text{IB}} \\ \hat{\underline{p}}_{\text{IB}} \end{pmatrix} + \begin{pmatrix} \underline{\underline{A}}_{NS} & \underline{\underline{A}}_{\text{IB}} \\ \underline{\underline{\mathcal{M}}}_{\text{IB}} \\ \underline{\underline{\Phi}}_{\text{IB}} \end{pmatrix} \begin{pmatrix} \hat{\underline{p}} \\ \hat{\underline{\tilde{T}}} \\ \hat{\underline{u}}_1 \\ \hat{\underline{u}}_2 \\ \hat{\underline{u}}_3 \\ \hat{\underline{u}}_{1,\text{IB}} \\ \hat{\underline{u}}_{2,\text{IB}} \\ \hat{\underline{u}}_{3,\text{IB}} \\ \hat{\underline{p}}_{\text{IB}} \end{pmatrix} = \begin{pmatrix} \underline{b}_{\text{p}} \\ \underline{b}_{\text{T}} \\ \underline{b}_{\text{u}_1} \\ \underline{b}_{\text{u}_2} \\ \underline{b}_{\text{u}_3} \\ \underline{b}_{\text{u}_1,\text{IB}} \\ \underline{b}_{\text{u}_2,\text{IB}} \\ \underline{b}_{\text{u}_3,\text{IB}} \\ \underline{b}_{\text{p},\text{IB}} \end{pmatrix} \qquad (13)$$

where $\underline{\underline{A}}_{NS}$ represents the TWE discretized in the interior of the domain and $\underline{\underline{\mathcal{M}}}_{\text{IB}}$ the frequency-transformed momentum equation extended to the center of each IB face, where pressure and velocity components are co-located. The latter contribute to the TWE in the interior of the domain via the block $\underline{\underline{A}}_{\text{IB}}$. While four new unknowns have been introduced at the IB, only three new equations, one for each component of the momentum, have been added in the iHS formulation. Hence, additional $n_{\text{IB}}$ linearly independent equations are required for closure, where $n_{\text{IB}}$ is the number of discrete faces at the IB. Such closure conditions, represented by the block $\underline{\underline{\Phi}}_{\text{IB}}$ are discussed in the following section.

### 2.4. Proposed algebraic closure of the inverse Helmholtz problem

Limited to the scope of the present manuscript, the following class of closure conditions are considered:

$$\Psi_m = \frac{\hat{p}_{\text{IB},m+1}}{\hat{p}_{\text{IB},m}}, \qquad \text{for} \qquad m = \{1, \ldots, n_{\text{IB}} - 1\} \tag{14}$$

where $\Psi_m$ is the ratio between complex pressure amplitudes taken at two different locations on the IB (figure 4). The condition (14) yields the following system of equations

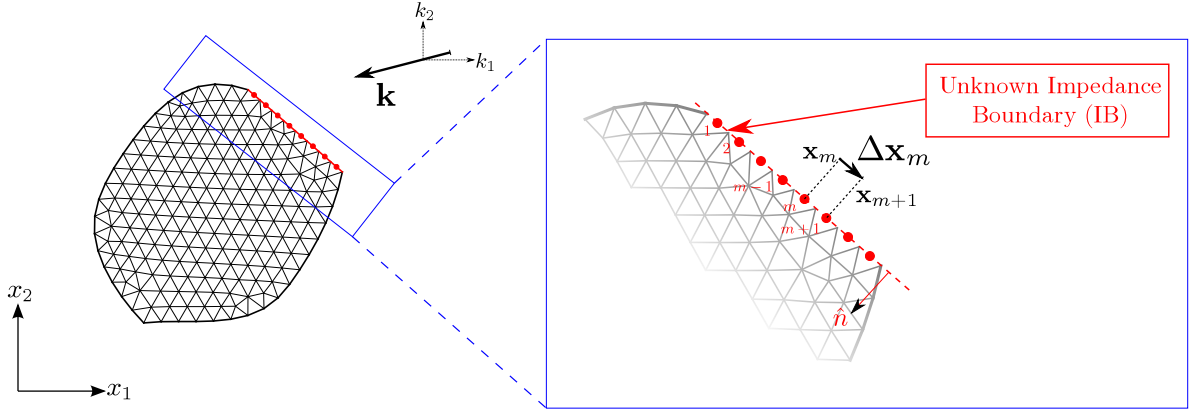

Figure 4: Discretization of IB surface and assignment of plane wave closure condition.

$$\begin{pmatrix} \Psi_1 & -1 & 0 & \cdots & 0 \\ 0 & \Psi_2 & -1 & \cdots & 0 \\ \vdots & \ddots & \ddots & \ddots & \vdots \\ \vdots & \ddots & \ddots & \Psi_{n_{\text{IB}}-1} & -1 \\ 0 & 0 & 0 & 0 & 0 \end{pmatrix} \begin{pmatrix} \hat{p}_{\text{IB},1} \\ \hat{p}_{\text{IB},2} \\ \vdots \\ \vdots \\ \hat{p}_{\text{IB},n_{\text{IB}}} \end{pmatrix} = \begin{pmatrix} 0 \\ 0 \\ \vdots \\ \vdots \\ 0 \end{pmatrix}$$

or,

$$\underline{\underline{\mathcal{P}}}_{\text{IB}} \cdot \underline{\hat{p}}_{\text{IB}} = \underline{0}. \tag{15}$$

For example, assuming harmonic planar wave propagation at the IB in the absence of a mean flow [1], the ratio (14) reads,

$$\Psi_m = e^{j \mathbf{k} \cdot \Delta \mathbf{x}_m}, \tag{16}$$

where, $\mathbf{k} = \{k_1, k_2, k_3\}$ is the wavenumber vector, and $\Delta \mathbf{x}_m = \mathbf{x}_{m+1} - \mathbf{x}_m$ is the displacement vector from the $m$-th to the $(m+1)$-th IB face (figure 4).

The remaining condition is the assignment of an arbitrary non-zero value of the complex pressure amplitude, hereafter $\hat{p}_{\text{ref}}$ (figure 3b), to an arbitrary internal cell; the resulting impedance at the IB is independent from $\hat{p}_{\text{ref}}$ due to the linearity of the TWE. This corresponds to the system of linear equations

$$\begin{pmatrix} 0 & \cdots & \cdots & \cdots & 0 \\ \vdots & \ddots & \ddots & \ddots & \vdots \\ 0 & \cdots & \cdots & \cdots & 1 \end{pmatrix} \begin{pmatrix} \hat{p}_1 \\ \hat{p}_2 \\ \vdots \\ \hat{p}_{n_{\text{cv}}} \end{pmatrix} = \begin{pmatrix} 0 \\ \vdots \\ \hat{p}_{\text{ref}} \end{pmatrix}$$

or,

$$\underline{\underline{\mathcal{P}}}_{\text{ref}} \cdot \underline{\hat{p}} = \underline{b}_{\text{p,IB}}, \tag{17}$$

where the $n_{\text{cv}}$-th cell has been arbitrarily chosen. Combining equations (15) and (17) yields

$$\underline{\underline{\mathcal{P}}}_{\text{ref}} \cdot \underline{\hat{p}} + \underline{\underline{\mathcal{P}}}_{\text{IB}} \cdot \underline{\hat{p}}_{\text{IB}} = \underline{b}_{\text{p,IB}}, \tag{18}$$

which forms the last row of the system (13), leading to,

$$j\omega\hat{\underline{p}}_{\text{IB}} + \underline{\underline{\Phi}}_{\text{IB}} \cdot \underline{X}_{\text{iHS}} = \underline{\underline{\mathcal{P}}}_{\text{ref}} \cdot \hat{\underline{p}} + \underline{\underline{\mathcal{P}}}_{\text{IB}} \cdot \hat{\underline{p}}_{\text{IB}}, \qquad (19)$$

where $\underline{X}_{\text{iHS}} = \left\{ \hat{\underline{p}}, \hat{\underline{T}}, \hat{\underline{u}}_1, \hat{\underline{u}}_2, \hat{\underline{u}}_3, \hat{\underline{u}}_{1,\text{IB}}, \hat{\underline{u}}_{2,\text{IB}}, \hat{\underline{u}}_{3,\text{IB}}, \hat{\underline{p}}_{\text{IB}} \right\}^T$.

Upon solving the iHS system (13), the impedance at the $m$-th IB face associated to the i-th velocity component can be directly evaluated as,

$$Z_{i,\text{IB}}|_m = \left( \frac{\hat{p}_{\text{IB}}}{\hat{u}_{i,\text{IB}}} \right)_m \qquad (20)$$

at any assigned angular frequency, $\omega$.

### 2.5. Considerations on acoustic power

The normal impedance, $Z_n$, can be evaluated as, $Z_{n,\text{IB}} = \hat{p}_{\text{IB}}/(\hat{\mathbf{n}} \cdot \mathbf{u})$, where $\hat{\mathbf{n}}$ is hereafter intended as the surface normal directed *into* the iHS domain. The normal resistance, $R_n$, explicitly appears in the isentropic expression of the cycle-averaged acoustic power, $d\dot{W}_{ac}$, transmitted through an infinitesimal area, $dA$, centered about a location $\mathbf{x}$ on a surface $\mathcal{S}$ of assigned impedance, $Z_n$. The surface power density, in fact, reads,

$$d\dot{W}_{ac}/dA = \frac{\omega}{2\pi} \int_0^{\frac{2\pi}{\omega}} p'(\mathbf{x};\tau) u_n'(\mathbf{x};\tau) d\tau = \frac{1}{2} \text{Re}\left[ \hat{p}^*(\mathbf{x};\omega) \hat{u}_n(\mathbf{x};\omega) \right] = \frac{1}{2} R_n(\mathbf{x};\omega) \hat{u}_n(\mathbf{x};\omega) \hat{u}_n^*(\mathbf{x};\omega) \qquad (21)$$

for a time-harmonic fluctuation of type (1). For a broadband periodic fluctuating field, the overall acoustic power can be obtained as the summation of tonal contributions (21) invoking orthogonality among different harmonics.

The result in (21) suggests that the reactance, $X(\mathbf{x};\omega)$ does not affect the acoustic power. This can be easily confuted with the following thought experiment. Let $\mathcal{S}$ be a planar surface of assigned normal resistance $\text{Re}\left[Z_n\right] = \rho_0\, a_0$. Assigning a zero reactance, $\text{Im}\left[Z_n\right] = 0$, allows a plane wave of (any) frequency $\overline{\omega}$ traveling along the direction of the normal $\hat{\mathbf{n}}$ to seamlessly cross the surface $\mathcal{S}$. The acoustic power transmitted through an infinitesimal surface element of $\mathcal{S}$ is $d\dot{W}_{ac} = \frac{1}{2} \rho_0\, a_0 \hat{u}_n(\mathbf{x};\omega)\, \hat{u}_n^*(\mathbf{x};\omega)\, dA$. On the other hand, assigning a reactance $\rho_0\, a_0/\text{Im}\left[X_n\right] = 0$ for $\omega \to \overline{\omega}$ to the same surface entails a complete reflection of the same wave and the surface acts as a purely reflective boundary at frequency $\overline{\omega}$. Consequently, the acoustic power transmitted through the surface is zero.

The normal resistance, $R_n$, also contains information regarding the direction of flow of acoustic energy. A positive normal resistance at a surface implies that acoustic power (if non-zero) is flowing *in the direction* of the chosen normal. A direct consequence of the choice of an inward directed normal (§2.4) is seen in the impedance analysis of thermoacoustically active cavities (§4) with a negative temperature gradient directed outward at the IB, resulting in a negative value for $R_n$.

### 2.6. Considerations on isentropic-wave assumptions (or lack thereof)

Spatially distributed broadband acoustic impedance (2) of a given surface defines the coupling between isentropic acoustic waves on either side of the surface, thus specifying only one degree of thermodynamic freedom. Under locally non-isentropic conditions, more information needs to be specified to account for the second degree of thermodynamic freedom, which could be provided, for example, in the form of the ratio of complex pressure to temperature or density amplitudes, the latter not necessarily equal to $a_0^2$ under such conditions. We note that no isentropic-flow approximations are made in the inverse Helmholtz Solver formulation (or when adopting the definition of acoustic impedance (2) itself) and the aforementioned ratios, despite not being analyzed in the present manuscript, are readily retrievable as an output of the solver.

## 3. Impedance of a thermoviscous straight duct

In this section validation of the iHS methodology is carried out in thermoviscous two-dimensional rectangular and circular ducts open on one end, with results compared against Rott's theory [14, 15] assuming quasi-one-dimensional wave propagation parallel to the duct axis (§3.1,3.2.1,3.2.3). Additionally, the normal and tangential impedance for different wave propagation angles at the open end are also evaluated (§3.2.2).

*3.1. One-dimensional iHS procedure applied to Rott's thermoviscous wave equations*

Rott's quasi-one-dimensional thermoviscous wave equations [14, 15],

$$j\omega\hat{p} = -\frac{1}{1+(\gamma-1)f_\kappa}\frac{\rho_0 a_0^2}{A}\frac{d\hat{U}}{dx}, \tag{22a}$$

$$j\omega\hat{U} = -(1-f_\nu)A\frac{1}{\rho_0}\frac{d\hat{p}}{dx}, \tag{22b}$$

discretized with a second order spatial scheme on a staggered grid (figure 5) become

$$j\omega\hat{p}_i = -\frac{1}{1+(\gamma-1)f_\kappa}\frac{\rho_0 a_0^2}{A}\frac{\hat{U}_{i+1}-\hat{U}_i}{\Delta x} \tag{23}$$

$$j\omega\hat{U}_{i+1} = -(1-f_\nu)A\frac{1}{\rho_0}\frac{\hat{p}_{i+1}-\hat{p}_i}{\Delta x}, \tag{24}$$

which can be spatially integrated for any given (real-valued) angular frequency $\omega$ from the hard end of the setup shown in figure 5 ($x=0$), starting with a given arbitrary reference pressure value, $\hat{p}_{\text{ref}}$, to the unknown impedance boundary at $x=L$, that is, with initial conditions

$$\hat{p}_1 = \hat{p}_{\text{ref}} \quad \text{and} \quad \hat{U}_1 = 0. \tag{25}$$

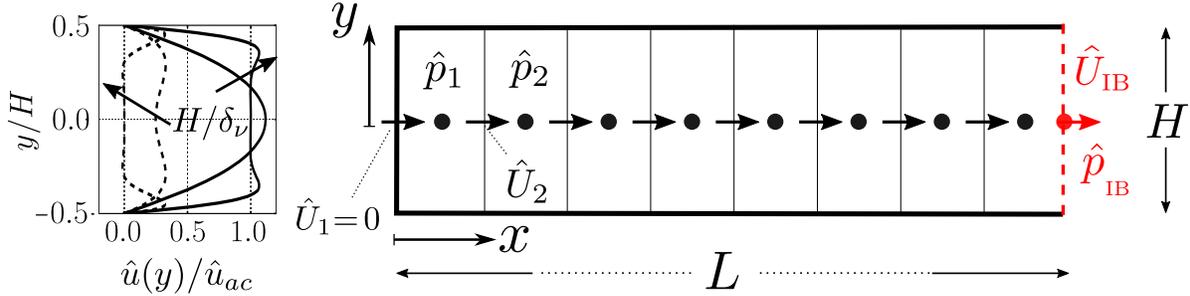

Figure 5: Staggered variable arrangement for one dimensional spatial integration of Rott's wave equations. The impedance at the end of the tube is calculated directly via (27). Dimensionless complex axial velocity profiles (26) normalized by isentropic velocity are shown for $H/2\delta_\nu = 4$ and $H/2\delta_\nu = 16$. Real part: ($--$); Imaginary part: ($—$).

In equations (22) and (24), $\hat{p}$ and $\hat{U}$ are the acoustic pressure and volumetric flow rate perturbations, and $f_\nu$ and $f_\kappa$ are the thermovisous functions [14],

$$f_\nu = \frac{(1+j)F_k'(\eta_0)}{\eta_0 F_k(\eta_0)}, \qquad f_\kappa = \frac{(1+j)F_k'\left(\eta_0\sqrt{Pr}\right)}{\eta_0\sqrt{Pr}F_k\left(\eta_0\sqrt{Pr}\right)}, \qquad \forall\ k=1,2$$

with $F_0(\eta) = \cosh(\eta), \quad F_1(\eta) = J_0(j\eta), \quad \eta_0 = y_0\frac{\sqrt{2j}}{\delta_\nu}, \quad \delta_\nu = \sqrt{\frac{2\nu}{\omega}},$ and $y_0 = \frac{H}{2},$

where $F_0$ and $F_1$ correspond respectively to two-dimesnsional cartesian and cylindrical duct geometries, $Pr = \mu c_p/\kappa$ is the Prandtl number, and $J_\nu$ are Bessel functions of the first kind of order $v$. The complex axial velocity at $x=L$ can be evaluated from the volumetric flow rate $\hat{U}$ such that [14],

$$\hat{u}(x=L,y) = \frac{\hat{U}_N}{A(1-f_\nu)}\left(1 - \frac{F_k(\eta)}{F_k(\eta_0)}\right), \text{ with } \eta = y\frac{\sqrt{2j}}{\delta_\nu} \text{ and } 0 \leq y \leq y_0, \tag{26}$$

thus allowing the calculation of the acoustic impedance at $x = L$ for each frequency, $\omega$, via,

$$Z(x = L, y; \omega) = \hat{p}_{\text{IB}}/\hat{u}_{\text{IB}}. \tag{27}$$

### 3.2. Two-dimensional thermoviscous duct

Figure 6 shows the two dimensional computational set up utilized for validation of the iHS methodology against Rott's linear theory for thermoviscous wave propagation (22) in rectangular ducts. For axisymmetric ducts, the iHS system of equations (13) was solved in cylindrical coordinates. Impedance was computed at the IB (figure 6) for an ideal and calorically perfect gas at homogeneous quiescent base-state,

$$u_0 = 0, \quad p_0 = 101325 \text{ Pa}, \quad \rho_0 = 1.2 \text{ kg/m}^3, \quad T_0 = 293.15 \text{ K}, \quad \gamma = 1.4, \text{ and} \quad a_0 = \sqrt{\gamma R T_0}.$$

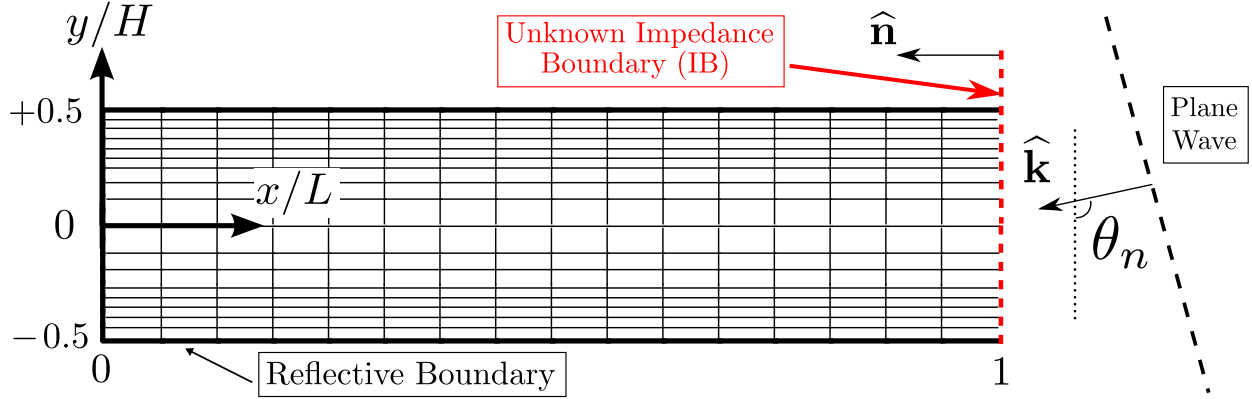

Figure 6: Computational set up for a two-dimensional duct with one open end analyzed via the inverse Helmholtz Solver (iHS). The grid shown here is just meant for illustrative purposes.

The present analysis is restricted to assumptions of planar wave propagation at the IB (16) in ducts of aspect ratio, A.R. $= L/H = 4$. For the setup shown in figure 6, equation (16) reduces to,

$$\Psi_m = e^{\text{j}(\mathbf{k}\cdot\Delta\mathbf{x})} = e^{\text{j}(k_x \Delta x + k_y \Delta y)} = e^{\text{j}(k \sin(\theta_n)\Delta x + k \cos(\theta_n)\Delta y)}. \tag{28}$$

In subsequent analysis, the frequency is expressed via the axial Helmholtz number, defined as,

$$He_L = \frac{fL}{a_0} = \frac{fH}{a_0}\text{A.R.} \tag{29}$$

where, $f = \omega/2\pi$ is the frequency of the planar wave.

#### 3.2.1. Axial wave propagation ($\theta_n = 90^o$)

The wave number vector, $\mathbf{k}$, in this case is perpendicular to the displacement vector, $\Delta\mathbf{x}_m$, (figure 4), yielding $\Psi_m = 1$ in equation (28).

Figure 7 shows the magnitude of specific admittance $|Y_{*,\text{IB}}|$ at the IB for an inviscid duct (rectangular or circular) evaluated using the iHS compared to the solution from Rott's inviscid (isentropic) wave propagation equations, that is, $f_\kappa = f_\nu = 0$ in equations (22) and (24). The regularly repeating admittance peaks occur at resonant frequencies, i.e. when the duct length is a multiple integer of the imposed wavelength, corresponding to the divergence of the surface normal admittance at the IB. At higher values of $He_L$, the iHS deviates from inviscid solution due to under-resolution at large wave-numbers. The value of impedance (and admittance) obtained from the iHS is purely imaginary for all inviscid test cases, consistent with the lack of thermoviscous wave attenuation effects [1]. The grid employed by the iHS runs in this case only has 3 cells in the transverse direction and 1000 cells in the axial direction.

For viscous test cases, 200 cells were utilized in the transverse direction with grid points concentrated near the isothermal no-slip boundaries ($y/H = \pm 0.5$ in figure 6) to resolve the thermoviscous boundary layers.

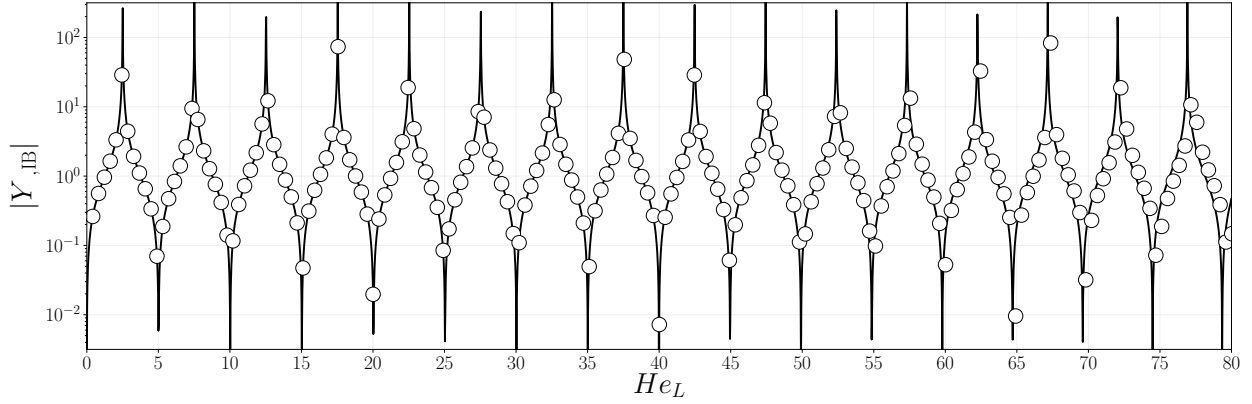

Figure 7: Magnitude of the purely imaginary surface averaged admittance obtained from the iHS (○) and from Rott's wave equations (—) for an inviscid two-dimensional duct.

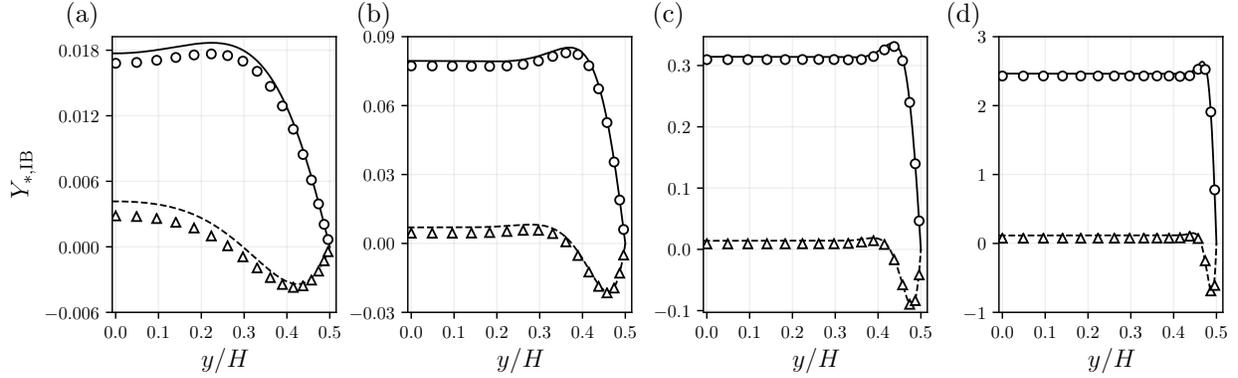

Figure 8: Comparisons of specific admittance obtained from the iHS (symbols) and Rott's theory (lines) at the IB of a two-dimensional viscous duct for dimensionless frequencies of $He_L = $ (a) $2.33 \times 10^{-3}$, (b) $1.163 \times 10^{-2}$, (c) $4.654 \times 10^{-2}$, and (d) $1.8614 \times 10^{-1}$. Real part: $(--, \triangle)$; Imaginary part: $(—, \circ)$.

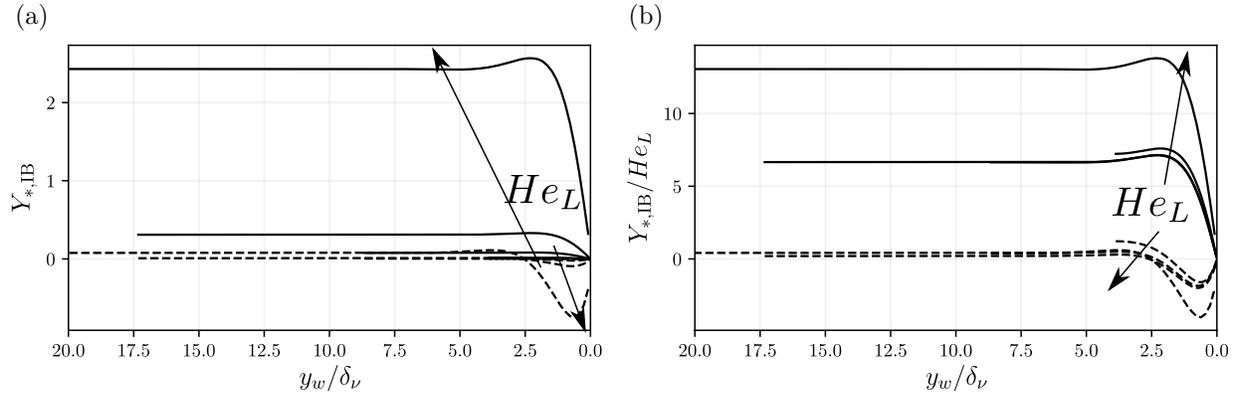

Figure 9: Real $(--)$ and imaginary $(—)$ components of specific acoustic admittance (a) and scaled by their respective Helmholtz numbers (b) plotted against distance from the wall, $y_w$ scaled by the corresponding Stokes' layer thickness, $\delta_\nu = \sqrt{2\nu/\omega}$.

At lower $He_L$, (figure 8a), the iHS results deviate from Rott's theory, where wall-normal pressure-gradients and velocity ($\partial \hat{p}/\partial y$ and $\hat{v}$) have been neglected due to an implicit assumption of very high duct aspect ratio. In reality, sufficiently close to the reflective end ($x=0$), both $\hat{u}$ and $\hat{v}$ are of comparable magnitude hence resulting in corner flow effects that are not accounted for by the theory. Such edge effects are, on the other hand, captured in the iHS runs yielding the correct results at low values of axial Helmholtz number. As $He_L$ increases (figure 8b-d), the Stokes' layer thickness is reduced (figure 11) as a result of which the flow affected by edge effects also shrinks thus resulting in better agreement.

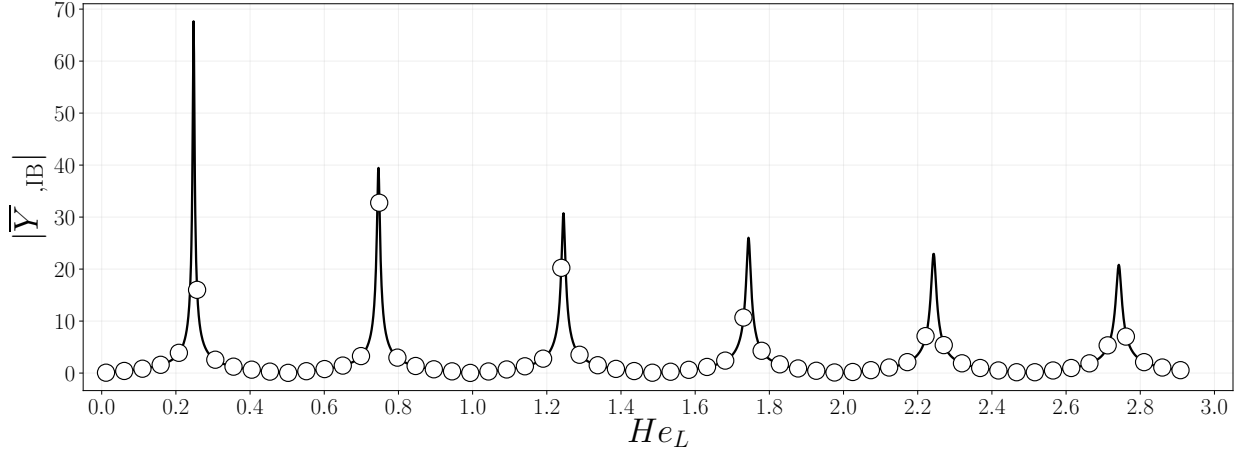

Figure 10: Magnitude of the surface averaged admittance obtained from the iHS (○) and from Rott's wave equations (—) for a two-dimensional rectangular viscous duct.

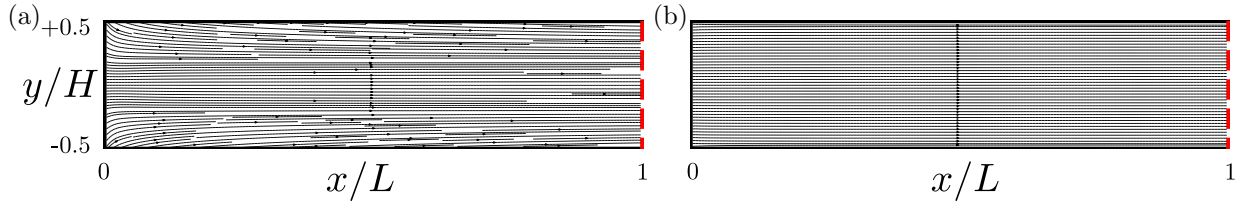

Figure 11: Streamlines inside the duct for axial wave propagation for frequencies, $He_L =$ (a) $2.33 \times 10^{-3}$, and (b) $1.8614 \times 10^{-1}$.

Figure 9a shows the admittance curves from figure 8 plotted against normal distance from the wall scaled by the corresponding Stokes' layer thickness, $\delta_\nu$. The location of the peaks in both real and imaginary curves appear to coincide on the $y_w/\delta_\nu$ axis. Additionally, scaling the admittance curves by their respective Helmholtz numbers, $He_L$, approximately collapses the curves for lower values of $He_L$.

Figure 10 compares the magnitude of the surface averaged admittance spectrum obtained from the iHS with semi-analytical solutions computed from Rott's wave equations up to a Helmholtz number, $He_L$ of 3. There is excellent overall agreement between the two results, especially beyond the threshold for diminishing edge effects as shown in figure 8d. Further, the peaks in admittance appear to be closer together when compared to the corresponding inviscid spectrum for the same duct (figure 7). Finally, the increasing gradient in $\left|\overline{Y}_{*,\text{IB}}\right|$ in the range $He_L = 1 \times 10^{-3}$ to $2 \times 10^{-1}$ that is considered in figure 8 does not allow the curves to perfectly collapse in figure 9b.

### 3.2.2. Obliquely planar wave propagation at IB

Figure 12 shows the surface admittance (normal and tangential) for incident angles in the range $\theta_n \in \left[\frac{\pi}{6}, \frac{\pi}{2}\right]$ (figure 6). Large values of $\text{Re}(Y_{*,IB})$ indicate high viscous losses near the walls. Moreover, since the frequency is smaller than the cut-off frequency, the wall normal (standing) modes are evanescent and decay exponentially along the duct. The tangential admittance (figure 12b) is evaluated along the direction of negative $y$ axis. In general, the peak magnitudes of real and imaginary components of tangential admittance are smaller than their normal admittance counterparts. Further, the peak magnitude of the real component of admittance decreases with increasing angle of incidence. At normal incidence, $\theta_n = \pi/2$, both real and imaginary components of tangential admittance vanish completely.

Figure 13 shows the variation of acoustic pressure fields (a) and streamlines (b) inside the duct for the various propagation angles considered. The setup is similar to that shown in figure 6 and the vertical red dashed lines at the right of every plot in figure 13 correspond to the IB. As the wave angle deviates from $\theta_n = \pi/2$, the streamlines near the IB start to curve, with some of them originating from and ending in the IB. This leads to a reversal in the sign of the real component of tangential admittance, $\text{Re}(Y_{*,t})$ (figure 12b), with the location of sign change being approximately the center of the streamlines' circulation region near the IB. Closer to the reflective end, the acoustic pressure and velocity fields appear to be unaffected by oblique incidence due to exponential decay of evanescent transverse modes.

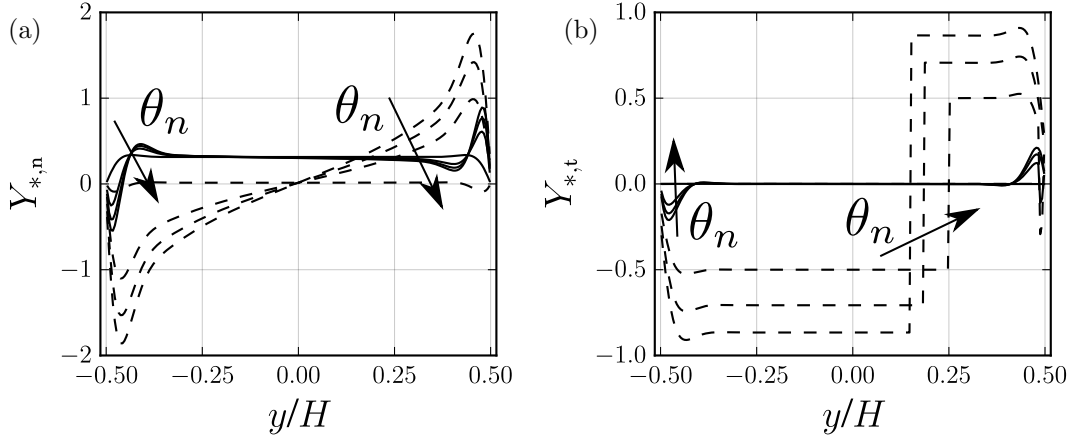

Figure 12: Spatial profiles of real (−−) and imaginary (—) components of normal (a) and tangential (b) specific acoustic admittance for viscous planar wave propagation at angles of incidence, $\theta_n = \frac{\pi}{6}, \frac{\pi}{4}, \frac{\pi}{3}$, and $\frac{\pi}{2}$.

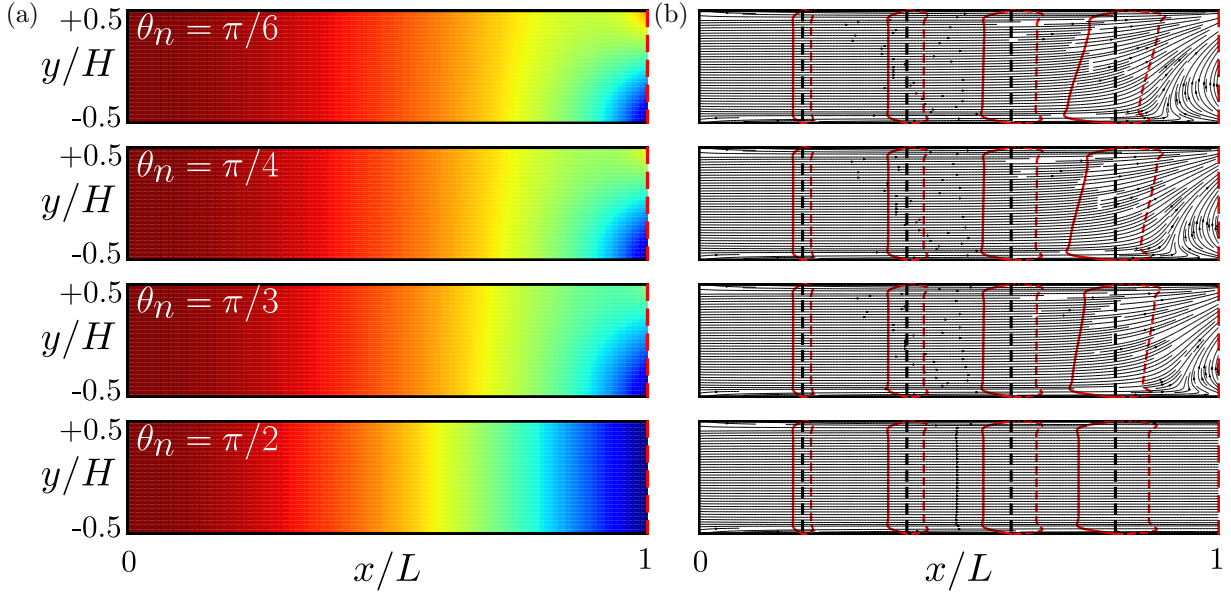

Figure 13: Acoustic pressure waveforms (a) and streamlines (b) inside the duct shown for wave propagation angles of $\theta_n = \frac{\pi}{6}, \frac{\pi}{4}, \frac{\pi}{3}$, and $\frac{\pi}{2}$ at the IB ($x = L$). The curved red lines (dashed and solid) represent real and imaginary components of axial velocity respectively at the axial locations shown by the vertical black dashed lines.

### 3.2.3. Circular cross-section duct

System (13) was recast in cylindrical coordinates to calculate the impedance at the open end of an axisymmetric duct. The aspect ratio, $AR$, in this case corresponds to the ratio of the height of the duct to its diameter. Figure 14 shows the comparisons of specific admittance, $Y_{*,\text{IB}}$ obtained from the iHS with semi-analytical solutions obtained from Rott's wave equations for an axisymmetric duct. Once again, at lower values of $He_L$ the iHS results deviate from Rott's theory, and this mismatch is minimized as the axial Helmholtz number increases, owing to diminishing edge effects as discussed in §3.2.1.

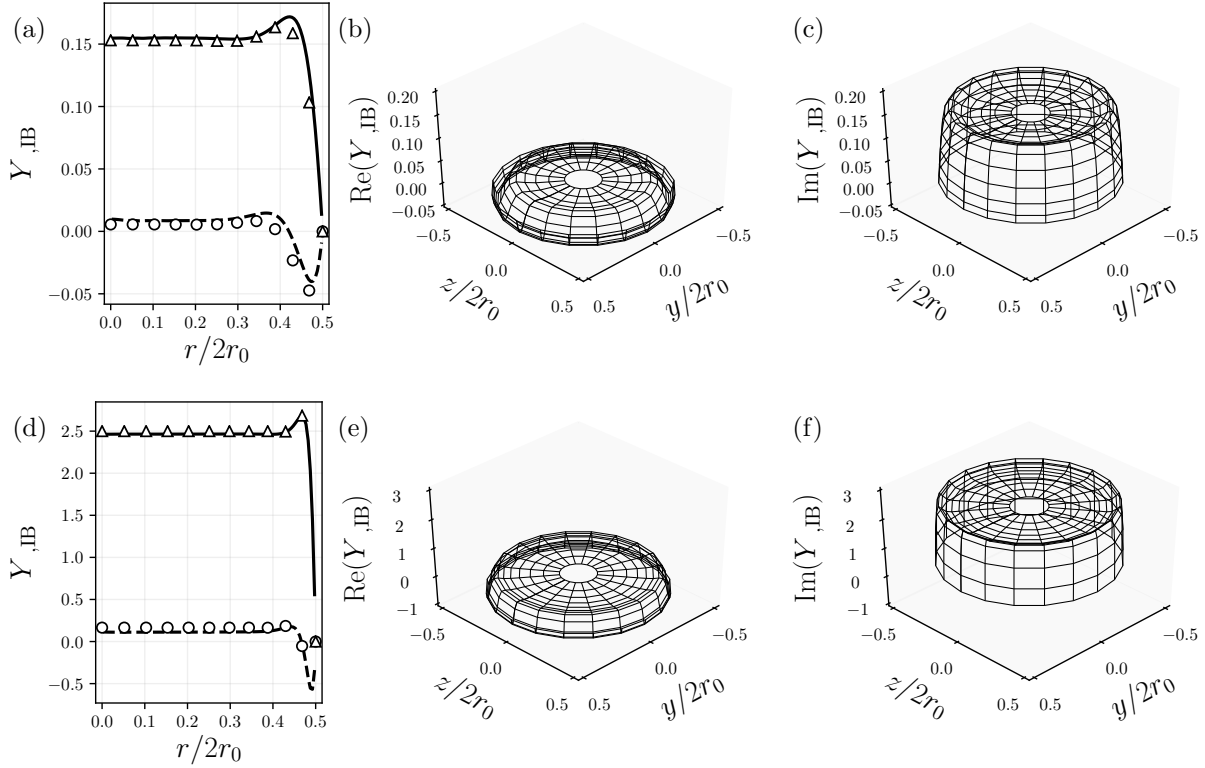

Figure 14: (a),(d) Comparisons of specific admittance obtained from the iHS (symbols) and Rott's theory (lines) at the IB of an axisymmetric viscous duct for dimensionless frequencies of $He_L =$ (a) $23.27 \times 10^{-3}$ and (d) $186.14 \times 10^{-3}$. Figures (b),(e) and (c),(f) portray respectively the real and imaginary parts of specific admittance obtained from the iHS in three dimensions. Real part: $(--,\triangle)$; Imaginary part: $(—, \circ)$.

## 4. Impedance of a thermoacoustically unstable cavity/pore

In this section, the iHS methodology is applied to a two-dimensional cavity with an imposed temperature gradient along the isothermal walls and in the base state. Figure 15 illustrates the computational set up utilized to extract the impedance at the unknown impedance boundary (IB) of the cavity (vertical red dashed line) from time-domain and cavity-resolved Navier-Stokes simulations. For the iHS, only the thermoacoustic cavity ($x \leq \ell_c$) needs to be considered. The temperature is varied linearly from $T_H$ to $T_C$ in the range $x = 0$ to $x = \ell_c$. The setup is a minimal-unit of a thermoacoustically unstable resonator. Initial conditions triggering first-mode standing wave instabilities are given by,

$$p(x, t=0) = p_0 + A \cos\left(\frac{x}{\ell_d + \ell_c}\pi\right), \quad u(x, t=0) = \frac{A}{Z_0}\sin\left(\frac{x}{\ell_d + \ell_c}\pi\right) \quad \text{with} \quad p_0 = 101325 \text{ Pa}, \quad (30)$$

where $\ell_d$ and $\ell_c$ are the axial lengths of the waveguide and the cavity, respectively and $A = 10$ Pa is the pressure amplitude. The base impedance, $Z_0 = \rho_0 a_0$ is evaluated using the ideal gas equation of state. Three different resonator (duct) lengths, $\ell_d$, are considered, resulting in three distinct resonant (unstable) frequencies. Key parameters and results corresponding to each test case are summarized in figure 15.

Figure 16 compares the spatial profiles of specific impedance (a,b,c) and admittance (d,e,f) obtained directly using the iHS with those extracted from the Navier Stokes simulations for the three different resonator lengths. The resistance, $\text{Re}(Z_*)$, in each case is a negative quantity. As discussed in §2.5, a negative value of normal resistance, $R_n$ for an inward directed surface normal implies production of acoustic energy in the cavity and acoustic flux directed away from the iHS domain. The thermoacoustic growth rates can, in fact, be quantified utilizing the impedance at the IB, as suggested by the expression (21). To this end, we utilize the cross-sectionally averaged thermoacoustic perturbation energy equation derived by Gupta *et al.* in [20], given by,

$$\frac{\partial E}{\partial t} + \frac{\partial I}{\partial x} = \mathcal{P} - \mathcal{D}, \tag{31}$$

where $\mathcal{P}$ and $\mathcal{D}$ denote thermoacoustic production and dissipation of energy and

$$E = \frac{1}{2}\rho_0 \left(\frac{U'}{w}\right)^2 + \frac{p'^2}{2\rho_0 a_0^2}, \text{ and } I = p'\frac{U'}{w}, \tag{32}$$

where $E$ is the acoustic energy density, $I$ is the acoustic flux, and $w = w_c$ or $w_d$. Outside the cavity, $\mathcal{P} = \mathcal{D} = 0$. Integrating from $x = \ell_c$ to $x = \ell_d + \ell_c$, the equation reads,

$$\frac{d}{dt}\int_{\ell_c}^{\ell_c+\ell_d} E \, dx = p' \left.\frac{U'}{w_c}\right|_{x=\ell_c}. \tag{33}$$

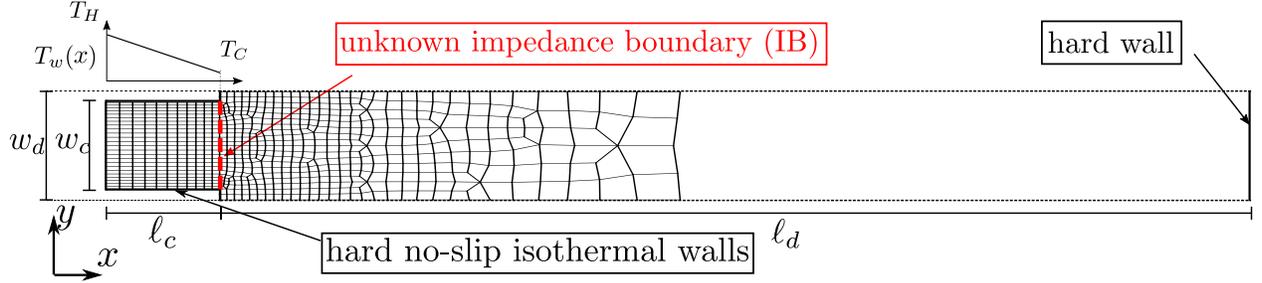

Figure 15: Computational set up (top) for impedance extraction of the thermoacoustically unstable cavity from pore-resolved Navier Stokes simulations. Test case parameters along with corresponding unstable mode frequencies, $\omega/2\pi$ and growth rates, $\alpha$ have been listed in the inset table. A constant dynamic viscosity, $\mu = 2.53 \times 10^{-5}$ kg/(m.s) was utilized in all calculations.

| $\ell_c$ (m) | $w_c$ (m) | $w_d$ (m) | $\ell_d$ (m) | $\omega/2\pi$ (Hz) | $\alpha$ (s$^{-1}$) | | | $T_H$ (K) | $T_C$ (K) |
|---|---|---|---|---|---|---|---|---|---|
| | | | | | LST | iHS | NS | | |
| $4 \times 10^{-3}$ | $10^{-4}$ | $1.25 \times 10^{-4}$ | 0.1 | 1682.95 | 7.03 | 6.98 | 8.14 | 600 | 300 |
| | | | 0.075 | 2220.12 | 14.34 | 14.81 | 17.82 | | |
| | | | 0.05 | 3258.32 | 32.37 | 36.49 | 34.06 | | |

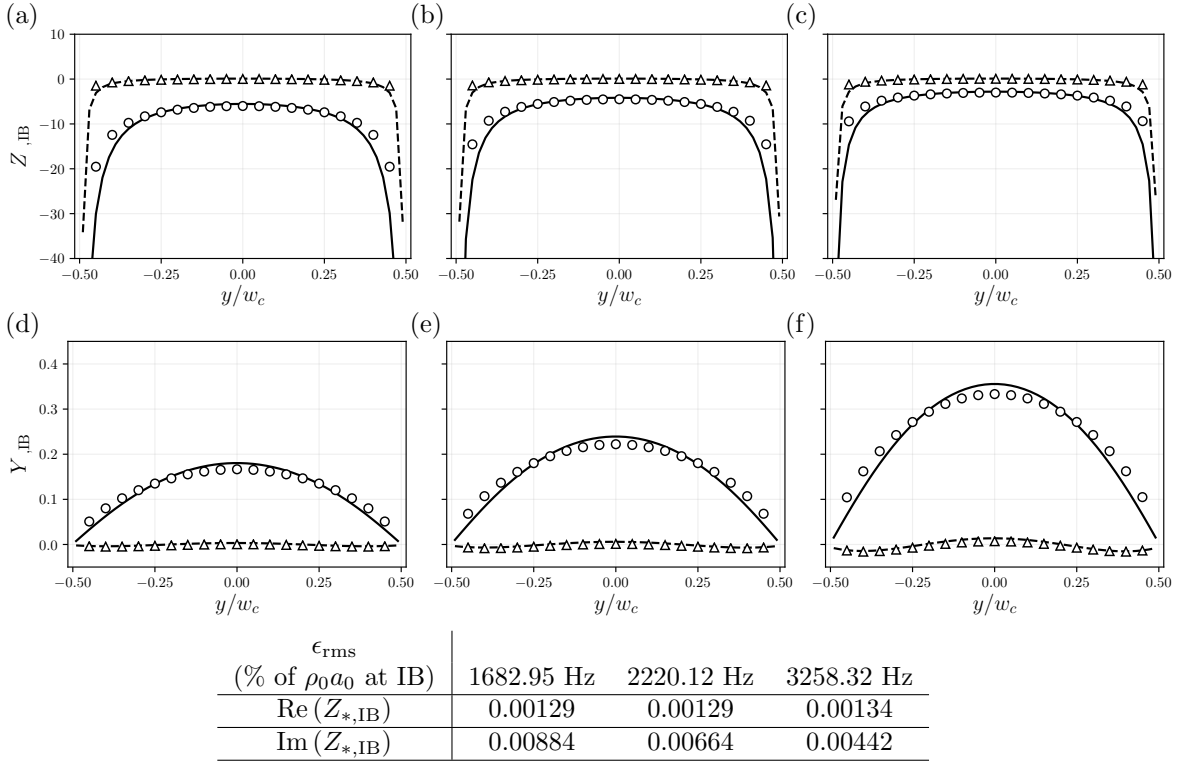

| $\epsilon_{\rm rms}$ (% of $\rho_0 a_0$ at IB) | 1682.95 Hz | 2220.12 Hz | 3258.32 Hz |
|---|---|---|---|
| $\text{Re}(Z_{*,\text{IB}})$ | 0.00129 | 0.00129 | 0.00134 |
| $\text{Im}(Z_{*,\text{IB}})$ | 0.00884 | 0.00664 | 0.00442 |

Figure 16: Specific acoustic impedance (top row) and admittance (bottom row) obtained from the iHS (lines) and extracted from fully compressible unstructured Navier Stokes simulations (symbols) of the thermoacoustic pore at the IB for frequencies, (a),(d) 1682.95 Hz; (b),(e) 2220.12 Hz; and (c),(f) 3258.32 Hz. Real part: $(--,\triangle)$; Imaginary part $(-,\circ)$.

Assuming modal growth of pressure and velocity perturbations, $\text{Re}\left(\hat{p}(x)e^{j\omega t}\right)$ and $\text{Re}\left(\hat{u}(x)e^{j\omega t}\right)$ (consequently $E \sim e^{2\alpha t}$), the growth rate, $\alpha$ can be evaluated as,

$$\alpha = \frac{1}{2E_d}\frac{dE_d}{dt} = \frac{\langle p'U'\rangle}{2w_c E_d} = -\frac{1}{2}\text{Re}(\overline{Y})\rho_0 a_0^2 \frac{|\hat{p}|^2_{\ell_c}}{\int_{\ell_c}^{\ell_c+\ell_d}|\hat{p}|^2(x)dx}, \tag{34}$$

where $\overline{Y}$ denotes cross-sectionally averaged admittance at the IB, $E_d = \int_{\ell_c}^{\ell_c+\ell_d}\langle E\rangle \, dx$, and $\langle \cdot \rangle$ denotes the cycle averaged quantities. The negative sign appears due to the orientation of the normal to the IB (figure 15). Assuming the waveform to be the first harmonic for an equal length inviscid duct, (30), the expression (34) can be approximated as,

$$\alpha \approx -\text{Re}(\overline{Y})\frac{\rho_0 a_0^2}{\ell_d}\cos^2\left(\frac{\pi\ell_c}{\ell_c+\ell_d}\right). \tag{35}$$

Predicted growth rates from the impedance obtained from inverse solver (iHS) and from Navier Stokes simulations (NS), using expression (35), are reported in figure 15 and compared with those obtained from linear stability theory (LST). Overall, there is good agreement in the predicted growth rates, and in the spatial profiles of impedance and admittance (figure 16) from the two solvers.

## 5. Impedance of a toy porous cavity

In this section a geometrically complex toy cavity is investigated using the iHS. In §5.2, impedance at the open end (IB) of the toy cavity obtained from the iHS is compared with that extracted from fully compressible Navier-Stokes simulation of a monochromatic wave incident normally on the cavity. Finally, application of the iHS methodology in the modeling of acoustic cavities as lumped impedance boundaries [16, 18] is discussed and demonstrated (§5.3). To this end, a broadband pulse reflection is studied, first off the pore resolved cavity, and then off an equivalent time domain impedance boundary condition (TDIBC) utilizing the iHS data, following the procedure in [17].

### 5.1. Cavity geometry and base state conditions

Geometry and a representational unstructured mesh of the analyzed two-dimensional toy cavity are shown on the left end in figure 17. Momentum and thermal perturbations vanish near the circular obstacles (no-slip isothermal boundaries) and oscillating boundary layers develop thus mandating appropriate near-wall resolution. For example, in calculations of hypersonic boundary layer transition control [21, 22], length scales of such cavities are very small compared to boundary layer thickness, making flow and pore-resolved calculations prohibitively expensive. However, impedance at the open end of the cavity (unknown impedance boundary, IB), if known, can be utilized in modeling the cavity as TDIBC in practical simulations thus eliminating the need for expensive pore-resolved simulations.

Test cases in the following subsections assume planar and axial wave propagation in a viscous medium at atmospheric base pressure and temperature $T_0 = 300$ K for air, modeled as an ideal gas. Dynamic viscosity is evaluated using Sutherland's law with an augmented reference viscosity, $\mu_{\text{ref}} = 1.827 \times 10^{-3}$ kg·m$^{-1}$·s$^{-1}$ to reduce grid resolution requirements.

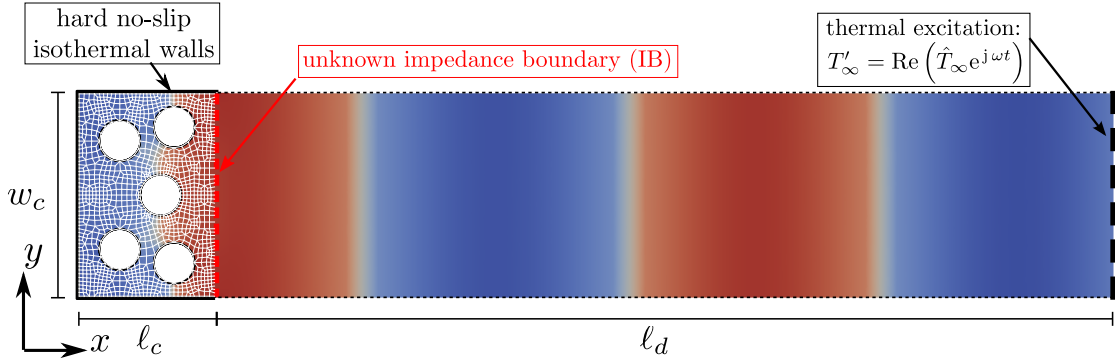

Figure 17: Computational set up for harmonic excitation and impedance extraction of the geometrically complex toy cavity, a representative mesh of which is shown in white. Pressure contours at steady state for $|\hat{T}_\infty| = 0.1$ K are also shown.

## 5.2. Harmonic excitation

Figure 17 shows the computational set up used for extracting the impedance at the mouth of the cavity corresponding to the unknown impedance boundary, IB (vertical red dashed line) from the steady state response of a harmonic (acoustic) excitation of the cavity. An isothermal boundary condition with harmonic variation in temperature was imposed at the right end and periodic boundary conditions were imposed on the lateral walls of the waveguide (dashed lines). At steady state, the acoustic pressure and velocity perturbations in the Fourier space $(\hat{p}, \hat{u})$ were extracted to evaluate the acoustic impedance. Due to the cavity geometry, pressure fluctuations are not perfectly planar at the IB. This allows us to apply and compare two different closure conditions for the iHS: (i) planar wave closure conditions as discussed in §2.4, and (ii) data-driven closure conditions derived from the pore-resolved Navier-Stokes simulations, which have been discussed below.

The application of a planar wave closure condition causes an expected mismatch in the spatial profiles of specific acoustic impedance (figure 18a-c) when compared to those extracted from pore-resolved simulations. Further, there is a sudden change in the sign of $\text{Re}(Z_{*,\text{IB}})$ close to the hard walls at low frequencies (figure 18a,b). These changes vanish, and the errors diminish, when acoustic pressure perturbations extracted from the pore-resolved simulations at the IB (figure 18d-f) are utilized as a closure condition with the iHS (figure 18g-i). Despite the mismatch, the overall errors are quite small (figure 18) and we adhere to planar wave propagation closure conditions in further analyses.

## 5.3. Broadband pulse excitation

Utilizing the iHS results, a broadband pulse reflection was studied with the cavity modeled as an equivalent TDIBC and compared with cavity resolved calculations. Initial conditions corresponding to a left traveling wave (figure 19) were utilized and are given by,

$$p' = A_0 \, e^{-\frac{5}{4}k^2(x-\bar{x})^2} \, \sin(2\pi kx), \quad v' = -\frac{p'}{\rho_0 \, a_0}, \quad \rho' = p'/a_0^2, \text{ and } T' = \frac{1}{\rho_0 R_{\text{gas}}} p' - \frac{T_0}{\rho_0} \rho', \qquad (36)$$

with $A_0 = 5$ Pa, $k = 200$ m$^{-1}$, and $\bar{x} = 0.03$ m.

For modeling the cavity as a TDIBC, surface averaged specific acoustic impedance from the iHS was fit to the three parameter (single oscillator) model [16, 18],

$$Z_* = R + \left(\omega X_{+1} - \omega^{-1} X_{-1}\right) \qquad (37)$$

in the frequency range 20–100 kHz (figure 20), bracketing the frequency content of the initial wave packet (36). The characteristic wavelength of the initial wave packet has been deliberately chosen to be larger than the cavity's dimensions to justify the assumption of planarity of the pressure field at the IB. As shown in §5.2, in the frequency range of $10 - 100$ kHz the planar wave propagation approximation holds with very small errors introduced in surface average impedance values (figure 18). Consequently, the pulse reflection in the present case is approximated as one dimensional.

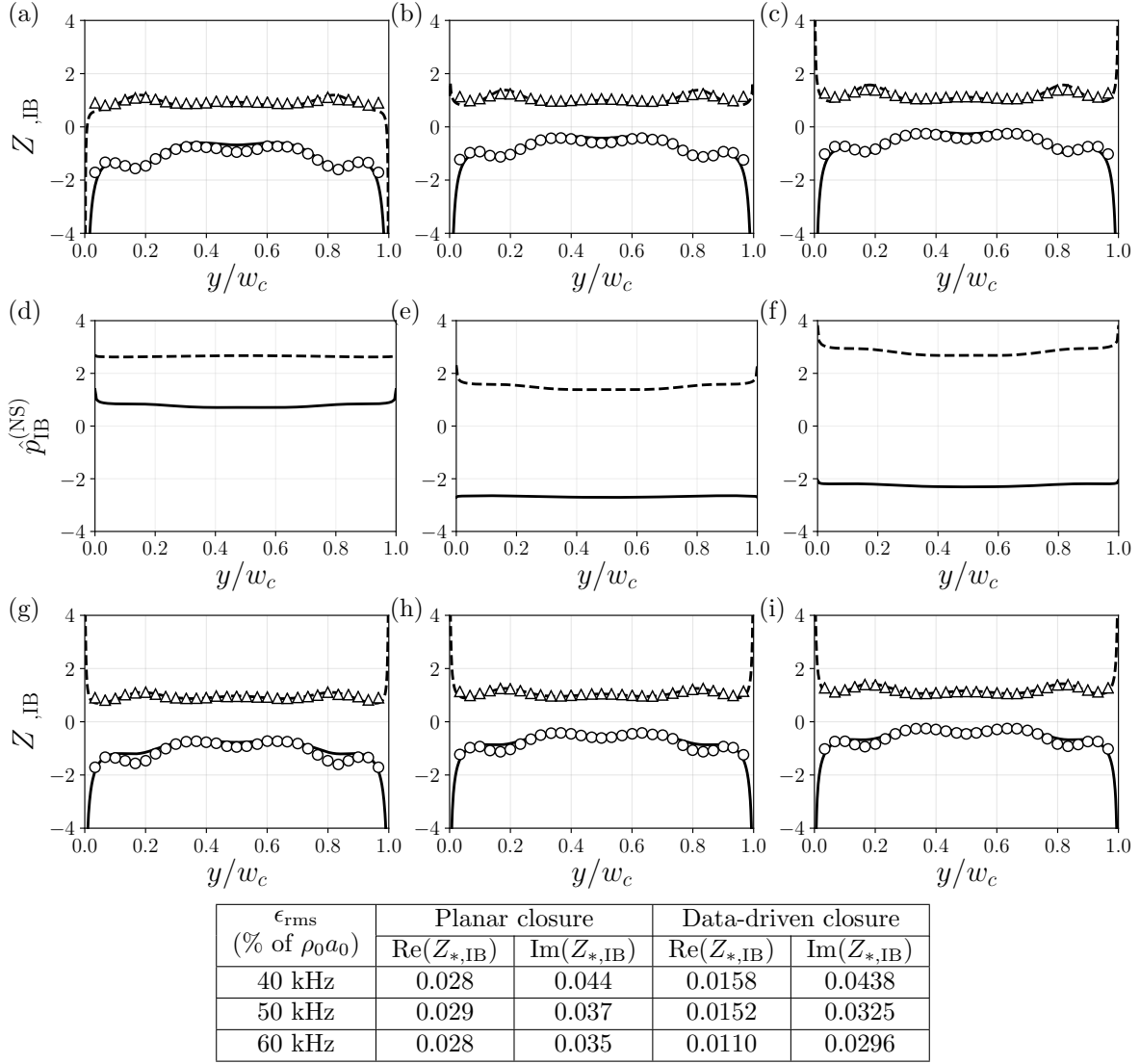

| $\epsilon_{\text{rms}}$ | Planar closure | | Data-driven closure | |
| :---: | :---: | :---: | :---: | :---: |
| (% of $\rho_0 a_0$) | $\text{Re}(Z_{*,\text{IB}})$ | $\text{Im}(Z_{*,\text{IB}})$ | $\text{Re}(Z_{*,\text{IB}})$ | $\text{Im}(Z_{*,\text{IB}})$ |
| 40 kHz | 0.028 | 0.044 | 0.0158 | 0.0438 |
| 50 kHz | 0.029 | 0.037 | 0.0152 | 0.0325 |
| 60 kHz | 0.028 | 0.035 | 0.0110 | 0.0296 |

Figure 18: Spatial profiles of specific acoustic impedance obtained from the iHS (lines) with a planar wave propagation assumption at the IB (a-c) and with data-driven closure closure conditions (g-i), compared with the impedance profiles extracted from pore-resolved Navier-Stokes simulations (symbols) for frequencies, (a),(d),(g) 40 kHz; (b),(e),(h) 50 kHz; and (c),(f),(i) 60 kHz. The extracted data-driven closure conditions for the three frequencies are also shown (d-f). Real part: $(--,\triangle)$; Imaginary part $(-,\circ)$.

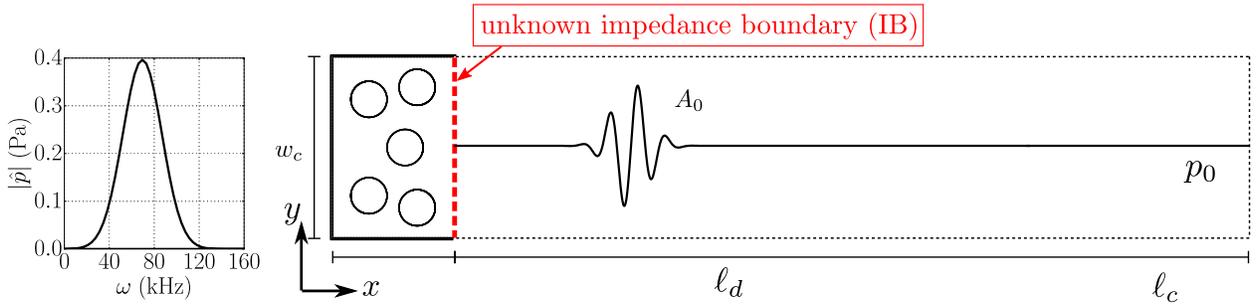

Figure 19: Initial setup (not to-scale) for the broadband pulse reflection off a toy cavity. Spectral content of the broadband acoustic pulse is shown on the left. Cavity dimensions: $l_d = 0.1$ m, $l_c = 0.001$ m, $w_c = 0.0015$ m.

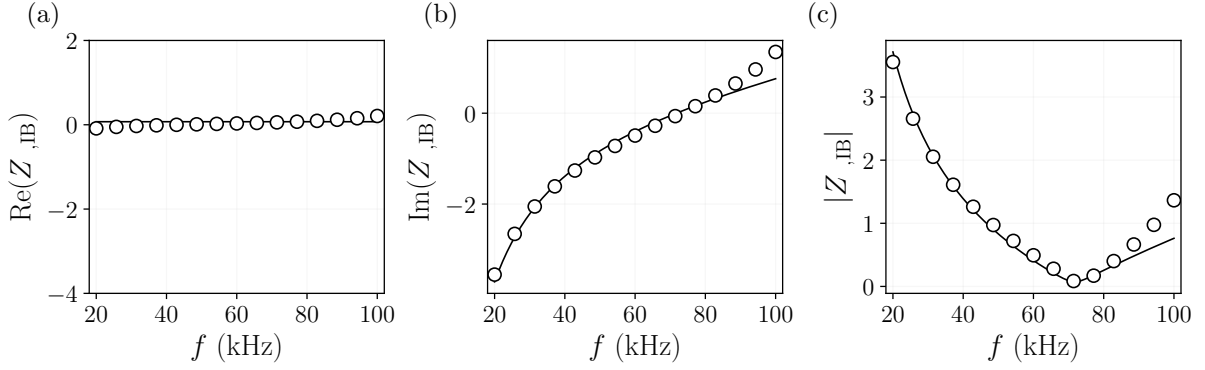

Figure 20: Comparisons between the (a) real part, (b) imaginary part, and (c) magnitude of the frequency-domain impedance obtained from the iHS (∘) and approximations using the three parameter model (—).

Figure 20 compares the results obtained from the iHS with the single oscillator fit that is obtained from the model (37), whose resonant frequency, $\omega_r = \sqrt{X_{-1}/X_{+1}}$, is approximately 73 kHz after the fit, representing the frequency of peak acoustic absorption of the cavity. Figure 20c shows that the fitted model is able to reproduce the cavity's absorption characteristics in the spectral neighborhood of the pulse (36). With the fitted TDIBC parameters, $R$, $X_{+1}$, and $X_{-1}$, time domain simulations following the procedure in Scalo *et al.* [17] have also been performed.

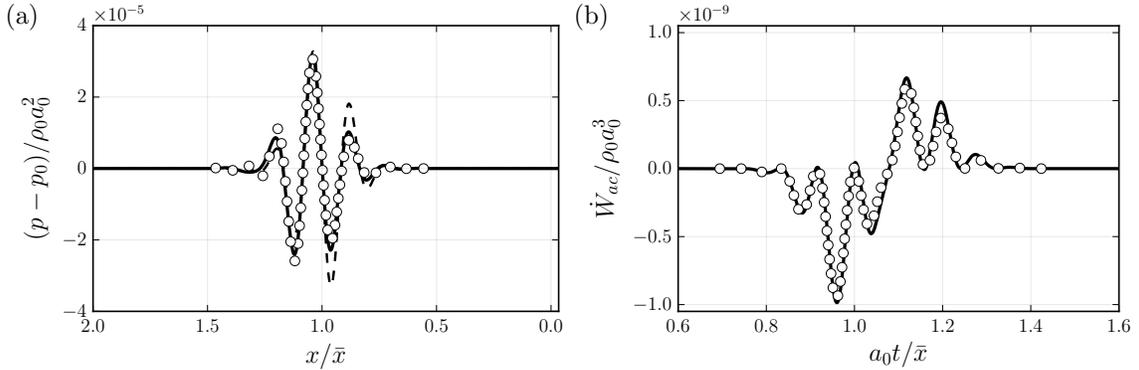

Figure 21: Comparison of TDIBC simulations (∘) with pore-resolved simulations (—): (a) spatial distribution of acoustic pressure after one bounce-back time, and (b) instantaneous power at the IB. The dashed lines in (a) correspond to the inital wave packet.

Figure 21a compares the spatial profiles of acoustic pressure from TDIBC and pore-resolved simulations after one reflection, i.e. at $t = 2\bar{x}/a_0$. Figure 21b compares the surface averaged instantaneous acoustic power, $\dot{W}_{ac}$, flowing through the IB (figure 17) in the two simulations. The latter shows that the total acoustic energy entering the duct (given by the area under the negative parts of the curve) is larger than the energy leaving the duct (given by the area under the positive parts of the curve), consistent with absorption of acoustic energy by the toy cavity. There is a small overall mismatch between the results from the two simulations, which can be attributed to the surface averaging of two-dimensional iHS data, and the use of a single-oscillator fit [16, 17, 18] as opposed to a more accurate multi-oscillator fit [23] in the evaluation of the TDIBC. The latter analysis will be deferred to future work.

## 6. Conclusion

We have presented an inverse Helmholtz Solver (iHS) methodology to determine the linear acoustic impedance at one (or multiple) unknown impedance boundaries (termed IB) of an arbitrarily shaped domain in response to acoustic wave propagation in the domain. We show that, for a given frequency, assignment of the spatial distribution of the pressure phase over the IB, which could physically correspond to the shape of the propagating wave, allows for mathematical closure of the problem (§2).

Utilizing the proposed methodology, we calculated the impedance at the open end of a two-dimensional rectangular and circular duct, assuming both viscous and inviscid wave propagation (§3). These results were compared—and the methodology validated—against those obtained from Rott's quasi one-dimensional linearized thermoviscous wave equations (22). In general, excellent matching was obtained for inviscid ducts, and for viscous ducts at high frequencies, owing to diminishing edge effects due to non-zero wall-normal gradients near the closed (reflective) end of the duct, which are neglected in Rott's theory. Furthermore, impedance of the duct for non-axial wave propagation at the IB is analyzed.

Following this, we analyzed a two-dimensional thermoacoustically unstable resonator with a cavity subject to a linear temperature gradient (§4) using the iHS. The spatial profiles of impedance reconstructed via the iHS at the mouth of the cavity match those obtained from fully resolved time domain simulations. A negative value for resistance, $\text{Re}(Z_{*,\text{IB}})$, obtained from both the iHS and from time domain simulations indicates that energy is produced in the cavity and flows out through the IB. Further, the growth rates obtained from the impedance extracted with both methods are consistent with results from the linear stability analysis.

Finally, we analyzed a geometrically-complex toy cavity using the iHS (§5). We first compared the specific acoustic impedance obtained from the iHS with that extracted from Navier Stokes simulations of harmonic excitations of the cavity. We then modeled the cavity as a lumped impedance boundary by transforming surface averaged impedance data from the iHS into a time domain impedance boundary condition (TDIBC) following Scalo *et al.* [17] and compared broadband pulse reflection simulations involving the TDIBC with equivalent simulations of reflection off the true pore-resolved cavity. In spite of various approximations employed in the formulation of the TDIBC, results from the two tests were found to be in good agreement, highlighting the effectiveness in recovering impedance data using the iHS.

The iHS methodology can serve as a straightforward, standalone tool to evaluate the broadband impedance of subdomains or components that could then be modeled as lumped impedance boundaries. One example is the modeling of microscale surface cavities encountered in hypersonic boundary layer transition control [21, 22] as an impedance surface, leaving only the flow side to be simulated. Future work involves applying the iHS to problems in hypersonics (boundary layer transition control), noise control (acoustic liners), and thermoacoustics (combustion).

### Acknowledgements

The authors acknowledge the support of the Rosen Center for Advanced Computing (RCAC) at Purdue and the Air Force Office of Scientific Research (AFOSR) award FA9550-16-1-0209. The author Carlo Scalo is thankful for the very fruitful discussions with Dr. Ivett Leyva (AFOSR). The authors would also like to thank Thomas Rothermel, Dr. Alex Wagner, and Dr. Markus Kuhn at the German Aerospace Center (DLR) for extending their support for hypersonic applications. Finally, the author Danish Patel would like to thank Mario Tindaro Migliorino for assisting in the derivation of the generalized equation of state that is currently implemented in the inverse Helmholtz Solver (iHS).